\documentstyle[preprint,aps]{revtex}
\newcommand{\mb}[1]{ {\mbox{\boldmath{$#1$}}}  }

\begin{document}
\tightenlines
\title{The importance of self-consistency in determining interface 
properties of $S-I-N$ and $D-I-N$ structures.} 

\author{A.M. Martin$^{\dagger}$ and  James F. Annett$^{\S}$}
\address{$^{\dagger}$D\'{e}partment de Physique Th\'{e}orique, 
Universit\'{e} de Gen\`{e}ve, 1211 Gen\'{e}ve 4, Switzerland.}
\address{$^{\S}$University of Bristol, H.H. Wills Physics Laboratory, 
Royal Fort, Tyndall Ave,
Bristol BS8 1TL, United Kingdom.}

\date{\today}
\maketitle

\begin{abstract}

We develop a method to solve the Bogoliubov de Gennes equation for
superconductors self-consistently, using the recursion method.  The
method allows the pairing interaction to be either local or non-local
corresponding to $s$ and $d$--wave superconductivity, respectively.
Using this method we examine the properties of various $S-I-N$ and $D-I-N$
interfaces.  In particular we self-consistently calculate the spatially 
varying density of states and the superconducting order parameter. We see 
that changing the strength of the insulating barrier, at the interface, does
not, in the case of an $s$--wave superconductor, dramatically, change 
the low energy local density of states, in the superconducting region 
near the interface. This is in stark contrast to what we see in the case of
a $D-I-N$ interface where the local particle density of states is changed 
dramatically. Hence we deduce that in calculating such properties as the 
conductance of $S-I-N$ and $D-I-N$ structures it is far more important
to carry out a self-consistent calculations in the $d$--wave case. 
\end{abstract}

\narrowtext
\section{Introduction.}
\label{introduction}
\setcounter{footnote}{1}

The Bogoliubov 
de Gennes equations and approximations to these equations, for example the 
Eilenberger equations have in the last 10 years been used to great sucsess
 to examine the properties of normal superconducting 
hybrid structures. Most of this work, in recent times, has centered upon
solving these equations for the electrical mesoscopic tranport properties 
of such hybrid structures for a recent review of calculations of electrical
conductance properties of normal superconducting hybrid structures see 
Lambert and Ramondi
\cite{Colin98}. Virtually all of 
this work has been centred around the properties of systems which contain 
conventional superconducting regions, ie. the superconducting gap has been 
isotropic. Also all but a few  
\cite{Bruder90,Hara93,Barash95,xiang,miller,Can95,Can97,Martin95,Martin96,Chang96,Riedel96} 
have been non-self-consistent 
calculations, this however has not inhibited the sucsess of such calculations. 
In more recent times, with the discovery of high-temperature superconductivity 
again the Bogoluibov de Gennes equations have been used to examine the
properties of hybrid structures, where now the superconducting gap is 
taken to be non-isotropic, usually $d$--wave in symmetry, but mainly so far 
most of the theoretical analysis of {\it simple} experiments 
\cite{wollman,sun,tsuei} have relied upon macroscopic symmetry arguments 
and not the microscopic
details of the actual interfaces.  However, the microscopic
physics at the interface can be an important factor in understanding
the experimental results. For example, if mixing of different
order parameter symmetries occurs at the interface (because the
interface breaks the bulk tetragonal or orthorhombic symmetry)
the extent of such mixing can only be determined from
microscopic calculations. Similarly, suppression of the order parameter
(either $d$--wave or $s$--wave) near an $S-N$ interface can lead to
significant local density of states within the bulk 
energy gap, and this can complicate the analysis of
single electron tunnelling spectra\cite{kitizawa}. 

In the past few years there have been a number of
microscopic calculations of surfaces and interfaces in
superconducting systems with a $d$--wave
order parameter. Most of the theoretical results have been obtained 
using tunnelling theory, or Andreev's 
approximation\cite{CRHu,XMT,TK,ZWT,YTanaka,SYMHK,TWZ,MMX,TK1,LHSNHYMK} 
in which the tunnelling
barrier and the order parameter are not found self-consistently.
For tunnel junctions these approximations may be adequate, but we show
below that self-consistency has significant effects for interfaces
with direct contact between the constituents. 
Self-consistent properties of interfaces have previously been computed
using the 
Eilenberger equations \cite{BGZ,MS1,MS2,MS3,BSB,MKM}, which are an 
approximation to
the Bogoliubov de Gennes equation. These self-consistent solutions to
the Eilenberger equations have shown some interesting effects.  Also there 
have now in the last couple of years 
been a few self-consistent calculations of the Bogoliubov de Gennes 
equations, with non-local interactions, looking at the effect of impurity 
scattering \cite{YYYK} and superconducting 
interfaces \cite{ref,Martin98}.

In this paper we tackle a {\it simple} problem, solving the Bogoliubov de
Gennes equations self-consistently, in the presence of an $s$--wave and a 
$d$--wave order parameter with the aim of asking, how important is 
self-consistency? We present  a method of how to calculate 
self-consistent properties of superconducting interfaces by directly
solving the Bogoliubov de Gennes equations \cite{Martin98} and that 
certainly in the case of a $d$--wave interface it is crucial to perform 
such a calculation to obtain the correct form for the low energy density 
of states in the region of an interface. 
The method of performing our self-consistent calculations will be
described in section II.
Here we introduce the Bogoliubov de Gennes equation 
with a general interaction, $U_{ij}$, and
demonstrate how this general Hamiltonian can be solved 
self-consistently using the Recursion
Method \cite{RHaydock,LMG,Martin98}.  We then present results for bulk 
$s$--wave and $d$--wave systems; looking at both the local particle density 
of states and in the case of $d$--wave we look at the critical temperature 
as a function of band filling, we see that a $d$--wave solution, for the 
interaction we have chosen, is not the only possible solution, for
our Hamiltonian.

Then in section III we present results for 
both $S-I-N$ and $D-I-N$ systems. We 
examine both the form of the superconducting order parameter in
the region of the interface, for different strengths of insulating 
barrier, and how the local particle density of states varies
in the region of the interface. What we see is the sub-gap 
local particle density of states, on the superconducting side of the 
interface, is not in the case of $s$--wave superconductivity effected 
by changing the strength of the insultaing barrier, this is in marked 
contrast to what we see in the $d$--wave case.
Finally in the conclusions we comment on what can be said in the light of 
our calculations.

\section{The Bogoliubov de Gennes Equation.}

Our starting point is the non-local Hubbard model, which is described by the 
following Hamiltonian
\begin{equation}
\label{eq:Hub}
H=-\sum_{i \, j \, \sigma} t_{ij} c^{\dagger}_{i \sigma} c_{j \sigma} - \mu
\sum_{i \, \sigma} c_{i \sigma}^{\dagger} c_{i \sigma} + \frac{1}{2}
\sum_{i \, j \, \sigma} U_{i j} c^{\dagger}_{i \sigma} c_{i \sigma} 
c^{\dagger}_{j -\sigma} c_{j -\sigma}
\end{equation}
where the creation and annihilation operators $c^{\dagger}_{i \sigma}$ and 
$c_{i \sigma}$, respectivily, create and annihilate electrons with
spin $\sigma$ in the orbital centred at the lattice point labelled by $i$, 
$t_{ij}$ is the amplitude for hopping from site $j$ to site $i$, $\mu$ is the
chemical potential and $U_{ij}$ is the interaction energy of two electrons
with opposite spin on sites $i$ and $j$. To obtain the, generalized non-local, 
Bogoliubov de Gennes Equation from equation (\ref{eq:Hub}) we first make a 
mean field approximation for the {\it pairing field}
\begin{equation}
\Delta_{ij}=-U_{ij}\langle c_{i \sigma} c_{j -\sigma} \rangle
\end{equation}
and assuming that the fluctuations about this mean are small we then 
perform the Bogoliubov cononical-transformation, enabling us to obtain
\begin{equation}
\sum_{j} \mb{H_{ij}} 
\left( \begin{array}{c}
u_{j}^{n} \\
v_{j}^{n}
\end{array}
\right)
=
E_{n}
\left(
\begin{array}{c}
u_{i}^{n}\\
v_{i}^{n}
\end{array}
\right) \label{eq:1}
\end{equation}
with

\begin{equation}
\label{eq:1a}
\mb{H_{ij}}= \left( \begin{array}{cc}
H_{ij} & \Delta_{ij}  \\
\Delta^{\star}_{ij} & 
-H^{\star}_{ij}
\end{array} \right)
\end{equation}
where $u_i^n$ and $v_i^n$ are the particle and hole amplitudes, on
site $i$, associated with an eigenenergy $E_n$ and
where $\Delta_{ij}$ is the, possibly non-local,
pairing potential or gap function, which does not have to be real.

In the fully self-consistent Bogoliubov de Gennes equation the
normal state Hamiltonian $H_{ij}$ is given by
\begin{equation}
\label{eq:2}
H_{ij}=(t_{ij}+\frac{1}{2} U_{ij} n_{ij}) +
(\epsilon_{i} - \mu)\delta_{ij} + \frac{1}{2}U_{ij} n_{jj}  
\end{equation}
where $\mu$ is the
chemical potential,  $\epsilon_{i}$ is the
normal on site energy of site $i$ and $t_{ij}$, as mentioned above is the 
hopping integral between site $i$ and site $j$. The interaction terms 
$\frac{1}{2} U_{ij} n_{jj}$ and $\frac{1}{2} U_{ij} n_{ij}$ are the 
Hartree-Fock potentials, which are defined by
\begin{equation}
\label{eq:3}
n_{ij}=\sum_{\sigma}\langle c_{i \sigma}^{\dagger} c_{j \sigma}\rangle=
2\sum_{n}((u_i^n)^{\star}u^n_j f(E_n)+v_i^n(v_j^n)^{\star}(1-f(E_n)).
\end{equation}
Similarly the pairing potentials are defined as

\begin{equation}
\label{eq:4}
\Delta_{ij}=-U_{ij}F_{ij}
\end{equation}
where the anomalous density is

\begin{equation}
\label{eq:5}
F_{ij}=\langle c_{i \sigma} c_{j -\sigma}\rangle=
\sum_{n}(u_i^n (v^n_j)^{\star}(1- f(E_n))
-(v_i^n)^{\star}(u_j^n)f(E_n).
\end{equation}
In equations (\ref{eq:3}) the sum only considers terms $E_n$ up to the 
condenstate chemical potential ($\mu$) as compared to equation (\ref{eq:5}) 
where the sum is over all terms.

A solution to the above system of equations will be fully self-consistent
provided that both the normal ($U_{ij}n_{ij}$) and anomalous 
($\Delta_{ij}$) potentials are determined consistently
with the corresponding densities
$n_{ij}$ and $F_{ij}$ via equations (\ref{eq:3})  and (\ref{eq:5}).

Figure 1 illustrates the geometry corresponding to this system of equations, 
on a tight binding lattice.  
The tight-binding lattice has nearest neighbour hopping interactions 
($t_{ij}$), as well as  a coupling between particle and hole space, via a 
superconducting order parameter ($\Delta_{ij}$). If the interactions are 
purely on-site ($U_{ii}$) attractions then the pairing potential will be 
purely local ($\Delta_{ii}$), corresponding to the dashed line in figure 1. 
On the other hand when the interaction is non-local ($U_{ij}$, $i \neq j$)
the pairing potential $\Delta_{ij}$ will also be non-local,
as illustrated, for nearest nieghbour interactions, by the solid lines in 
figure 1.

\subsection{The Recursion Method for solving the Bogoliubov de Gennes 
Equations.}

If we now wish to solve the Bogoliubov de Gennes equations self-consistnetly, 
we have found that the most efficient method of doing this is
the Recursion Method \cite{RHaydock}.
This method allows us to calculate the electronic Green's functions

\begin{equation}
\label{eq:6}
G_{\alpha \, \alpha^{\prime}}(i,j,E)=
\langle i \alpha|\frac{1}{E \mb{1}- \mb{H}} |j \alpha^{\prime}\rangle
\end{equation}
where the indices $i$ and $j$ denote sites,
while   $\alpha$ and $\alpha^{\prime}$ represent the 
particle or hole
degree of
freedom on each site. We denote particle degrees of freedom by
$\alpha=+$ and hole degrees of freedom by $\alpha=-$. For example
$G_{+ \, -}(i,j,E)$ represents the Greens function between the
particle degree of freedom on site $i$ and the hole degree of freedom
on site $j$.

To compute the Green's functions (\ref{eq:6}) we can closely follow the 
method described by
Litak, Miller and Gy\"orffy\cite{LMG}, they solved the Bogoliubov de 
Gennes Equation self-consistently, using the recursion method, for a local 
interaction. Using this method we can transform the Hamiltonian to a block
tridiagonal form

\begin{equation}
E \mb{1}- \mb{H}=
\left(\begin{array}{cccccccc}
E\mb{1}-\mb{a_0} & -\mb{b_1} & 0 & 0 & 0 & 0 & 0 & \cdots \\
-\mb{b^{\dagger}_1} & E\mb{1}-\mb{a_1} & -\mb{b_2} & 0 & 0 & 0 & 0 & \cdots \\
0 & -\mb{b^{\dagger}_2} & \ddots & \ddots & 0 & 0 & 0 &\cdots \\
0 & 0 & \ddots & \ddots & \ddots & 0 & 0 & \cdots \\
0 & 0 & 0 & -\mb{b^{\dagger}_n} & E\mb{1}-\mb{a_n} & -\mb{b_{n+1}} & 0 &
\cdots \\
\vdots & \vdots & \vdots & 0  & \ddots & \ddots & \ddots & \ddots 
\end{array}
\right)
\label{eq:7}
\end{equation}
where $\mb{a_n}$ and $\mb{b_n}$ are $2 \times 2$ matrices.  Given this
form for $\langle i \alpha|E \mb{1}- \mb{H}|j \alpha^{\prime}\rangle$
and expressing the Green's function as

\begin{equation}
G_{\alpha \, \alpha^{\prime}}(i,j,E)=
\langle i \alpha|(E \mb{1}- \mb{H})^{-1} |j \alpha^{\prime}\rangle,
\label{eq:8}
\end{equation}
the Greens functions above can be evaluated as a matrix continued
fraction so that

\begin{equation} 
\mb{G}(i,j,E)= 
\left(E\mb{1}-\mb{a_{0}}-\mb{b^{\dagger}_1} 
\left(E\mb{1}-\mb{a_{1}}-\mb{b^{\dagger}_2} 
\left(E\mb{1}-\mb{a_{2}}-\mb{b^{\dagger}_3} 
\left( E\mb{1}-\mb{a_{3}}-\ldots \right) ^{-1} \mb{b_3} 
\right) ^{-1}\mb{b_2} \right) ^{-1}\mb{b_1} \right) ^{-1} 
\label{eq:9}
\end{equation}
where

\begin{equation}
\mb{G}(i,j,E)=\left(
\begin{array}{cc}
G_{\alpha \, \alpha}(i,i,E) & G_{\alpha \, \alpha^{\prime}}(i,j,E) \\
G_{\alpha^{\prime} \, \alpha}(j,i,E) & 
G_{\alpha^{\prime} \, \alpha^{\prime}}(j,j,E)
\end{array}
\right).
\label{eq:10}
\end{equation}

Within equations (\ref{eq:7}) and (\ref{eq:9}) we have a formally exact
representation of the Green's functions.  However in general 
both the tridiagonal
representation of the Hamiltonian, and the 
matrix continued fraction 
(\ref{eq:7}) will be infinite.  In practice one can only calculate
a finite number of terms in the continued fraction exactly.
In the terminology of the recursion method
it is necessary to {\em terminate} the continued fraction
\cite{RHaydock,LMG,AMagnus,DTT,CMMNex,GAllan,TKW}.

If we were to calculate up to
and including $\mb{a_n}$ and $\mb{b_{n}}$ and then simply set subsequent
coefficients to zero
then the Green's function
would have $2n$ poles along the real axis.  The density of states
would then correspond to a set of $2n$ delta functions.
Integrated quantities such as the the densities
$n_{ij}$ and $F_{ij}$ could depend strongly on $n$, especially
since only a few of the $2n$ delta functions would be within the relevant
energy range within the  BCS cut off, $E_c$. In order to 
obtain accurate results it would be necessary to calculate a large number
of exact levels, which would be expensive in terms of both computer
time and memory.

As a more efficient alternative we choose to terminate the continued
fraction using the extrapolation method, as used previously
by Litak, Miller and Gy\"orffy\cite{LMG}.
We calculate the values for
$\mb{a_n}$ and $\mb{b_{n}}$ exactly up to the first $m$ coefficients
using the recursion method.  Then, noting the fact that the elements of
the matrices
$\mb{a_n}$ and $\mb{b_{n}}$ vary in a predictable manner \cite{LMG}, we
extrapolate the elements of the matrices for a further $k$ iterations,
where $k$ is usually very much greater than $m$.  This enables
us to compute the various densities of states, and the
charge densities $n_{ij}$ and $F_{ij}$ accurately
with relatively little computer time and memory.

In terms of the Green's functions $G_{\alpha \,
\alpha^{\prime}}(i,j,E)$ the pairing and normal Hartree-Fock
potentials  
$\Delta_{ij}$ and $\frac{1}{2}U_{ij}n_{ij}$ can be expressed as

\begin{equation}
\label{eq:12}
\Delta_{ij}=\frac{1}{2 \pi} U_{ij}
\int (G_{+ \, -}(i,j,E+\imath \eta) - 
G_{+ \, -}(i,j,E-\imath \eta))(1-2f(E))dE
\end{equation}
and

\begin{equation}
\frac{1}{2}U_{ij}n_{ij}=\frac{1}{2\pi} U_{ij}
\int (G_{+ \, +}(i,j,E+\imath \eta) - 
G_{+ \, +}(i,j,E-\imath \eta))f(E)dE
\label{eq:13}
\end{equation}
where $\eta$ is a small positive number.
 
To obtain the above equations we have used the fact that

\begin{equation}
\left(
\begin{array}{c}
u_i^n \\
v^n_i
\end{array}
\right)
\, \, \, \, {\rm and} \, \, \, \,
\left(
\begin{array}{c}
-(v_i^n)^{\star} \\
(u^n_i)^{\star}
\end{array}
\right)
\end{equation}
are the eigenvectors of equation (\ref{eq:1}) with eigenvalues $E_n$
and $-E_n$. 

It should be noted that up to this point we have been completely 
general, so we are not just restricted to calculating the Greens 
functions on the same site, $G_{\alpha \alpha^{\prime}}(i,i,E)$ or
just between nearest neighbours, in essence we could use this 
formalism to calculate a Greens function between any two sites.
This enables us to also calculate such quantities as 
$G_{++}(i,j,E)$ and $G_{+-}(i,j,E)$, which can then in turn 
be used to calculate normal reflection and 
transmission coefficients, 
thus enabaling in principle for transport properties or our 
Hamiltonian to be calculated.

\subsection{Achieving Self-consistency.}

Using the above methods to calculate $\Delta_{ij}$ and $U_{ij}n_{ij}$ 
we need to achieve a fully self-consistent solution.
Firstly we make use of any symmetries in the system
in order to minimise the number of calculations which are necessary.
For example on an infinite square lattice with no variation in
any of the potentials only one independent site needs to be calculated 
since this
site can be mapped onto all of the other sites.  Secondly,
once we have decided
which sites need to be calculated self-consistently $\Delta_{ij}$ and 
$n_{ij}$ can be calculated for those sites, remembering that
on a square lattice each site will have four nearest neighbours.
This implies that in general
 we will have to calculate  nine different Green's functions
in order to calculate $\Delta_{ij}$ and $n_{ij}$. This can be seen by
considering site $i$ in figure 1 and noting that we need to calculate
the Greens functions shown in table I, depending on whether the interaction 
is purely local, purely non-local, or both local and non-local.
Having calculated the appropriate Green's functions new
values for $\Delta_{ij}$ and $n_{ij}$ can be calculated, which we 
will denote as
$\Delta^{(1)}_{ij}$ and $n^{(1)}_{ij}$.
Inserting these into the Hamiltonian and repeating the calculation of then
Green's functions leads to a new set
$\Delta^{(2)}_{ij}$ and $n^{(2)}_{ij}$  and so on.
 We repeat this iteration for all $i$ and $j$ until

\begin{equation}
\label{eq:14}
\left|\frac{|\Delta^{(n-1)}|-|\Delta^{(n)}|}{|\Delta^{(n)}|}\right| \le 0.001
\end{equation}
and 

\begin{equation}
\label{eq:15}
\left|\frac{|n^{(n-1)}|-|n^{(n)}|}{|n^{(n)}|}\right| \le 0.001.
\end{equation}

Since $\Delta_{ij}$ and $n_{ij}$ can be complex we need to also check
for convergence in their associated phases.

\subsection{Uniform properties of our Hamiltonian.}

Up to this point we have said nothing about the interaction itself, 
just that it exists. In conventional BCS theory \cite{BCS} we can impose 
a cut-off 
for our interaction, since we know that in this case the interaction is
phonon  mediated, thus in equations (14) and (16) we can have a cut-off 
for our integrals. However we also wish to consider the properties of
high-temperature superconductors, where the mechanism for the interaction is
a contriversial area, so we state at this point that we are not interested in
the mechanism for high-temperature superconductivity, but we do know, through
experiment \cite{annett}, that the symmetry of the superconducting order
parameter in high-temperature materials is of a $d$--wave nature, therefore,
generally we wish to consider the propeties of $d$--wave superconductors.

The simplest model to consider is one where we have a nearest neighbour 
attractive interaction, $U_{ij}$. For such a situation, where the 
interaction is constant we can calculate the critical temperature 
for a given interaction strength, as a function of band filling, 
see figure 2. It is found that there
are two possible solutions for the symmetry of the superconducting 
order parameter. In this figure we can see there are two critical 
temperatures, one for extended $s$--wave solutions, near the edges 
of the band, and a $d$--wave solution near the centre of the band. 
It is this $d$--wave solution which we will be considering, however it is
important to note, and will be seen later, that in non-uniform systems 
although we may be at a band filling where only the $d$--wave solution 
exists this does not nessesitate that the extended $s$--wave components 
to the  superconducting order parameter will thence be zero.

Before proceeding it is instructive to examine the bulk local particle
density of states ($N(E)$) for the systems we are concerned with. In figure 
3 we have plotted the local particle density of states,

\begin{equation}
\label{eq:16}
N_i(E)=\frac{1}{2\pi} (G_{+ \, +}(i,i,E+\imath \eta)-G_{+ \,
+}(i,i,E-\imath \eta)).
\end{equation}
for three different scenarios, no interaction (thin dashed line), a local
interaction (thin solid line) and a, nearest nieghbour, non-local 
interaction (thick solid line). What we see from this is exactly what 
one would expect, i.e. for no interaction, we obtain the density of states 
for a regular, nearest nieghbour tight binding
lattice, for a local interaction we obtain a BCS type gap centred 
around the Fermi Energy, and finally for a non-local interaction
we see a typical $d$--wave gap, ie. the local particle density of states
goes to zero linearly at the Fermi energy. Having said that there is no way, 
at this point to descern that we actually have a $d$--wave solution, but 
we define a $d$--wave order parameter in the following manner  

\begin{equation}
|\Delta^{(d)}_i|=\frac{1}{4}|\Delta_{ij_{1}}-\Delta_{ij_{2}}+\Delta_{ij_{3}}-
\Delta_{ij_{4}}|,
\end{equation}
where $j_1$, $j_2$, $j_3$ and $j_4$ are the nearest nieghbours to $i$, moving 
in a clockwise or anticlockwise direction, this quantity is then the 
$d$--wave component to the order parameter at $i$, similarly we can also 
write 

\begin{equation}
|\Delta^{(s)}_i|=\frac{1}{4}|\Delta_{ij_{1}}+\Delta_{ij_{2}}+
\Delta_{ij_{3}}+\Delta_{ij_{4}}|
\end{equation}
for the extended $s$--wave component to the order parameter. For the 
results obtained in figure 3 $|\Delta^{(d)}_i|=0.38*t$ and 
$|\Delta^{(s)}_i|=0$.

\section{Results for $S-I-N$ and $D-I-N$ interfaces}

Now we can proceed to perform self-consistent calculations for interfaces.
Our {\it model} is as follows, we consider a system which does not
vary in the $y$ coordinate and has a step like function in the 
interaction strength in the $x$ coordinate; ie. for $x<0$ the interaction
is finite; whereas for $x\ge 0$ our interaction (be it local or 
nearest nieghbour non-local) is zero. The interface itself is modelled 
by a line of sites at $x=0$, where we change the on-site energy
of these sites in accordance with the strength of the barrier at the 
interface we wish to model. We then proceed to calculate the 
Bogoliubov de Gennes equations in the self-consistent manner prescribed 
above.

First we look at the results obtained for a local interaction 
($U_{ii}=-1.5t$). We see in figure 4 the nomalized magnitude 
of the superconducting order parameter (upper graph) and 
$\frac{1}{2}U(x)n(x)$ (lower graph), where in reference to the prevoius 
equations we have made a transformation from sites $i$ to position $x$. In 
this figure we see how these two quantities vary in the region of the 
interface ($x=0$), for different strengths of the barrier at the interface,
$\epsilon(0)=0$ (thin solid line),  $\epsilon(0)=0.5t$ (thin dashed line),
$\epsilon(0)=1.0t$ (thick solid line) and $\epsilon(0)=2.0t$ 
(thick dashed line). The main point to deduce from this is that the 
barrier has very little effect upon the actual order parameter or the 
local density. This is in comparison to what we see for an equivalent  
calculation for a non-local interaction, the results of which are shown in 
figure 5, where we have plotted the nomalised $d$--wave contribution to the 
superconducting order parameter (upper graph) and the extended $s$--wave 
contribution to the superconducting order parameter (lower graph), for 
the same interface configurations as used to obtain figure 4. What we 
see here is that although $\Delta^{(d)}(x)$ is not seroiusly modified in 
the presence of a barrier, $\Delta^{(s)}(x)$ is. We can see that as we 
increase the strength of the barrier $\Delta^{(s)}(x)$ in the region 
of the interface is reduced.

We now wish to focus our attention on the local particle density of states, 
near the interface. How this quantity changes will be important in 
deducing whether it is crucial to perform self-consistent calculations to 
obtain quantities which are reliant upon the local particle density of states 
near the interface. As an example we are going to focus on the sub-gap 
density of states and how this changes for different barrier 
strengths, and whether it is then valid to use non-self-consistent 
calculations in calculating such quantities as the sub-gap conductance?   

With this in mind we have in figure 6 plotted the low energy local particle 
density of states for a local interaction ($U_{ii}=-1.5t$), at $x=-2$. We have 
plotted this quantity for different strengths of barrier, 
$\epsilon(0)=0$ (thin dashed line), $\epsilon(0)=0.5t$ (thin solid line),
$\epsilon(0)=1.0t$ (thick dashed line), $\epsilon(0)=2.0t$ (thick solid line) 
and $\epsilon(0)=3.0t$ (thin dashed-dotted line).
Again, as in figure 4 there is little change in the quantity 
we are interested in. In fact from this plot we can say that a quantity such 
as the sub-gap, $S-I-N$ conductance depends very little on changes in 
the density of states and is totally governed by the tunnelling 
conductance of the barrier.

However again taking the $d$--wave case we can see an altogther 
different story. In figure 7 we have plotted equivalent local  
particle density of states as in figure 6, but for a $d$--wave interface. 
What we see is that the low energy density of states in the region of the 
barrier is modified significantly, when we change the strength of the barrier, 
and hence we can no longer state that the conductance of an $D-I-N$
interface will be mainly governed by the tunnelling conductance of the barrier.
This implies that in such calculations, when comparing, for example the
conductances of different $D-I-N$ interfaces, the details of the
barrier effect the sub-gap density of states on the superconducting side of 
the barier and hence self-consistency has an important role to play.  

Finally to highlight the point that the low energy local particle density of 
states
in the region of an interface is modified significantly, by the strength of 
the barrier at the interface, when the supercunducting region has a $d$--wave 
symmetry, we have in figure 8 plotted the local particle 
density of states at $x=-1$ (a), $x=-2$ (b), $x=-3$ (c) and $x=-4$ (d). for 
two different strengths of barrier $\epsilon(0)=0$ (thin solid line) and
$\epsilon=3.0t$ (thick solid line). From this figure we can see the 
{\it dramatic} change in the sub-gap local density of states. 
Particularly in figure 8(a) (nearest the interface) we see that in the
region of $E=0$, when thier is no interface barrier, there is always a 
significant density of states at the interface, as compared to 
$N(0) \approx 0$ 
when we have a finite barrier. 

\section{Concluding remarks.} 

In this paper we have developed the Recursion Method for calculating the 
Bogoliubov de Gennes equations self-consistently with both local and 
non-local interactions. Then as a {\it simple} problem we examined the
self-consistent local particle density of states in the region of both 
$S-I-N$ and $D-I-N$ interfaces, in particular we have focused on how 
the local particle density of states is modified as we change the 
strength of the barrier. What we see is that in the case of the $S-I-N$ 
interface the low energy particle density of states, on the superconducting 
side of the interface, does not change dramatically as we change the strength 
of the insulating barrier. This is in stark contrast to the comparitively 
dramatic change we see when we perform the same calculation for a $D-I-N$ 
interface. 

These results imply that in the case of $d$--wave interfaces the details of 
the the interface are important and hence to get these details correct we 
should perform a self-consistent calculation. If we do a non-self-consistent 
calculation for the case of the $d$--wave interface, i.e. just let the order 
parameter vary in a step like manner at the interface, then we see that the 
low energy density of states on the superconducting side of the barrier is 
unaltered as we change the strength of the insulating barrier, at the 
interface. This discrepancy, between self-consistent and non-self-consistent 
calculations, arises from the fact that we
have a mixing of symmetries in the region of the 
self-consistent $d$--wave interface and when a self-consistent calculation is 
performed we see that this mixing changes as the interface is changed, this 
effect will not be seen in a non-self-consistent calculation. 

These results may well explain why non-self-consistent
conductance calculations for hybrid $s$--wave structures have over the past 
years been so sucsessful, i.e the problem can simply be seen as calculating 
the tunnel conductance of the barrier, but also raises a warning flag that 
such a bold approach for $d$--wave structures may not be as effective, since 
our calculations show that the low energy density of states changes as we 
change the properties of the interface, we believe that the only way 
to obtain the correct low energy density of states behavoir  is to perform a 
self-consistent calculation.

\section{Acknowledgments.}

This work was supported by the EPSRC under grant number GR/L22454 
and by the TMR network Dynamics of Nanostructures.
We would like to thank B.L Gy\"orffy for useful discussions.

\begin{figure}
\caption{This is a schematic diagram of a tight binding lattice, 
with particle and hole degrees of freedom, $\Delta_{ij}$ couples particles 
on site $i$ to holes on site $j$. The difference between 
local and non-local pairing is highlighted by the dashed (local pairing) and 
solid (non-local pairing) lines.}
\end{figure}

\begin{figure}
\caption{The critical temperature vs. band filling for a fixed 
non-local, nearest nieghbour interaction, the dashed line is the
extended $s$--wave solution to the linearized gap equation and 
the solid line is the corresponding $d$--wave solution.}
\end{figure}

\begin{figure}
\caption{The local particle density of states for a $2$-D tight binding 
lattice, with no interactions ($U_{ij}=0$) (thin dashed line), with a 
local interaction ($U_{ii}=1.5t$) (thin solid line) and finally 
with a non-local nearest nieghbour interation ($U_{ij}=1.5t$) 
(thick solid line). For this system $t_{ij}=1$ for nearest neighbours only, 
$\mu=0$ and $\epsilon_i=0$.}
\end{figure}

\begin{figure}
\caption{The magnitude of the superconducting order parameter (upper graph), 
normalised to the magnitude at $x=-\infty$ and $1/2 U(x)n(x)$ (lower graph), 
as a function of $x$, where we have an interface at $x=0$, Each of the lines 
on the upper and lower graphs are for different strengths of barrier at $x=0$,
$\epsilon(0)=0$ (thin solid line),  $\epsilon(0)=0.5t$ (thin dashed line),
$\epsilon(0)=1.0t$ (thick solid line) and $\epsilon(0)=2.0t$ 
(thick dashed line).}
\end{figure}

\begin{figure} 
\caption{The magnitude of the $d$--wave component to the supercunducting 
order parameter nomalized to its value at $x=-\infty$ (upper graph) and the 
$s$--wave component to the superconducting order parameter (lower graph) as 
a function of $x$, for different strengths of barrier at $x=0$. We have
$\epsilon(0)=0$ (thin solid line),  $\epsilon(0)=0.5t$ (thin dashed line),
$\epsilon(0)=1.0t$ (thick solid line) and $\epsilon(0)=2.0t$ 
(thick dashed line).}
\end{figure}

\begin{figure}
\caption{The local particle density of states, at $x=-2$, for an attractive 
local interaction, for $x<0$. The different lines are again for different 
strengths of interaction, $\epsilon(0)=0$ (thin dashed line),  
$\epsilon(0)=0.5t$ (thin solid line), $\epsilon(0)=1.0t$ (thick dashed line),
 $\epsilon(0)=2.0t$ (thick solid line) and $\epsilon(0)=3.0t$ 
 (thin dashed-dotted line).}
\end{figure}

\begin{figure}
\caption{The local particle density of states, at $x=-2$, for an attractive 
non-local interaction, for $x<0$. The different lines are again for different 
strengths of interaction, $\epsilon(0)=0$ (thin dashed line),  
$\epsilon(0)=0.5t$ (thin solid line), $\epsilon(0)=1.0t$ (thick dashed line),
 and $\epsilon(0)=2.0t$ (thick solid line) 
 (thin dashed-dotted line).}
\end{figure}

\begin{figure}
\caption{The local particle density of states, at $x=-1$ (a), $x=-2$ (b), 
$x=-3$ (c) and $x=-4$ (d), for an attractive 
non-local interaction, for $x<0$. The different lines are again for different 
strengths of interaction, $\epsilon(0)=0$ (thin solid line), and 
$\epsilon(0)=3.0t$ (thick solid line).}
\end{figure}

\begin{table}
\begin{tabular}{|c||c|c|c|c|c|}
\hline
Interaction Type & 
$\mb{G_{+ -}}(i, i,E )$ & 
$\mb{G_{+ \pm}}(i, j_1,E)$  & 
$\mb{G_{+\pm}} (i, j_2,E)$ & 
$\mb{G_{+\pm}}(i, j_3,E)$ & 
$\mb{G_{+\pm}}(i, j_4,E)$ \\
\hline
$U_{ii}$ & Y & N & N & N & N   \\
\hline
$U_{ij}(1-\delta_{ij})$ & N & Y & Y & Y & Y  \\
\hline
$U_{ii}+U_{ij}(1-\delta_{ij})$ & Y & Y & Y & Y & Y \\
\hline
\end{tabular}
\caption{This table shows which Greens functions need to be calculated 
for systems with interactions which are
local, $U_{ii}$, non-local, $U_{ij}(1-\delta-{ij})$,
or both.  The site labels correspond to the notation of Fig. 1.}
\end{table}


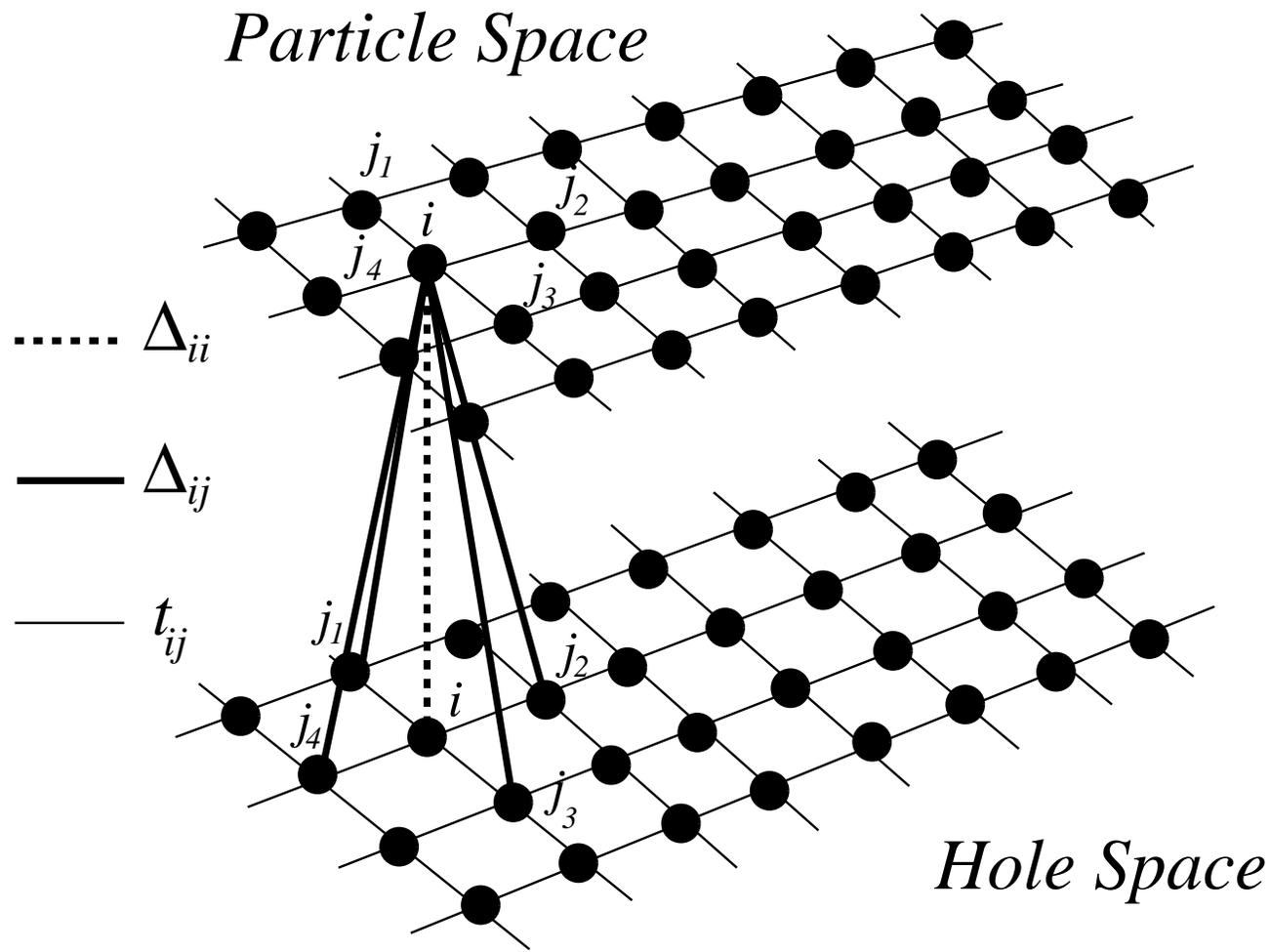
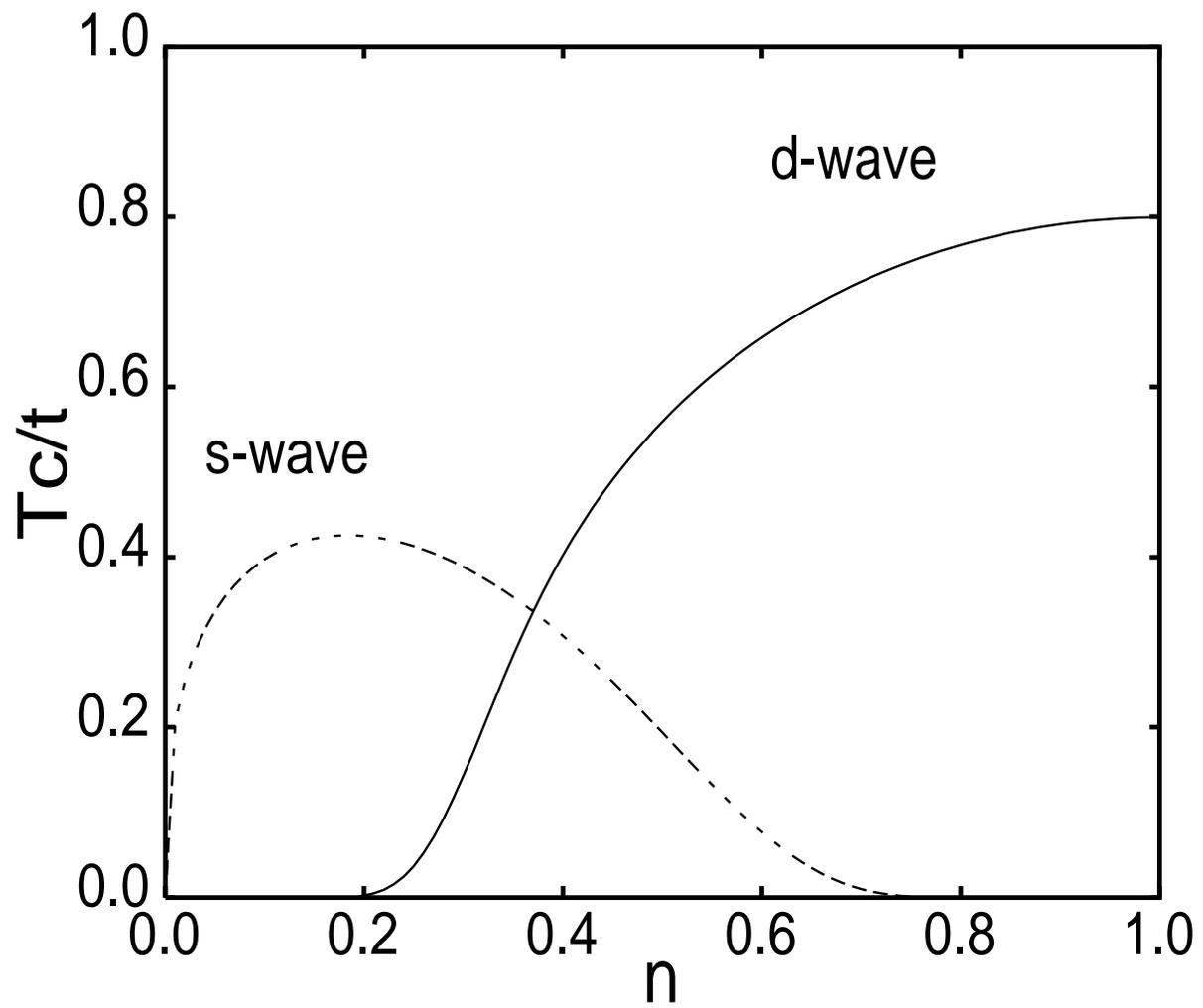
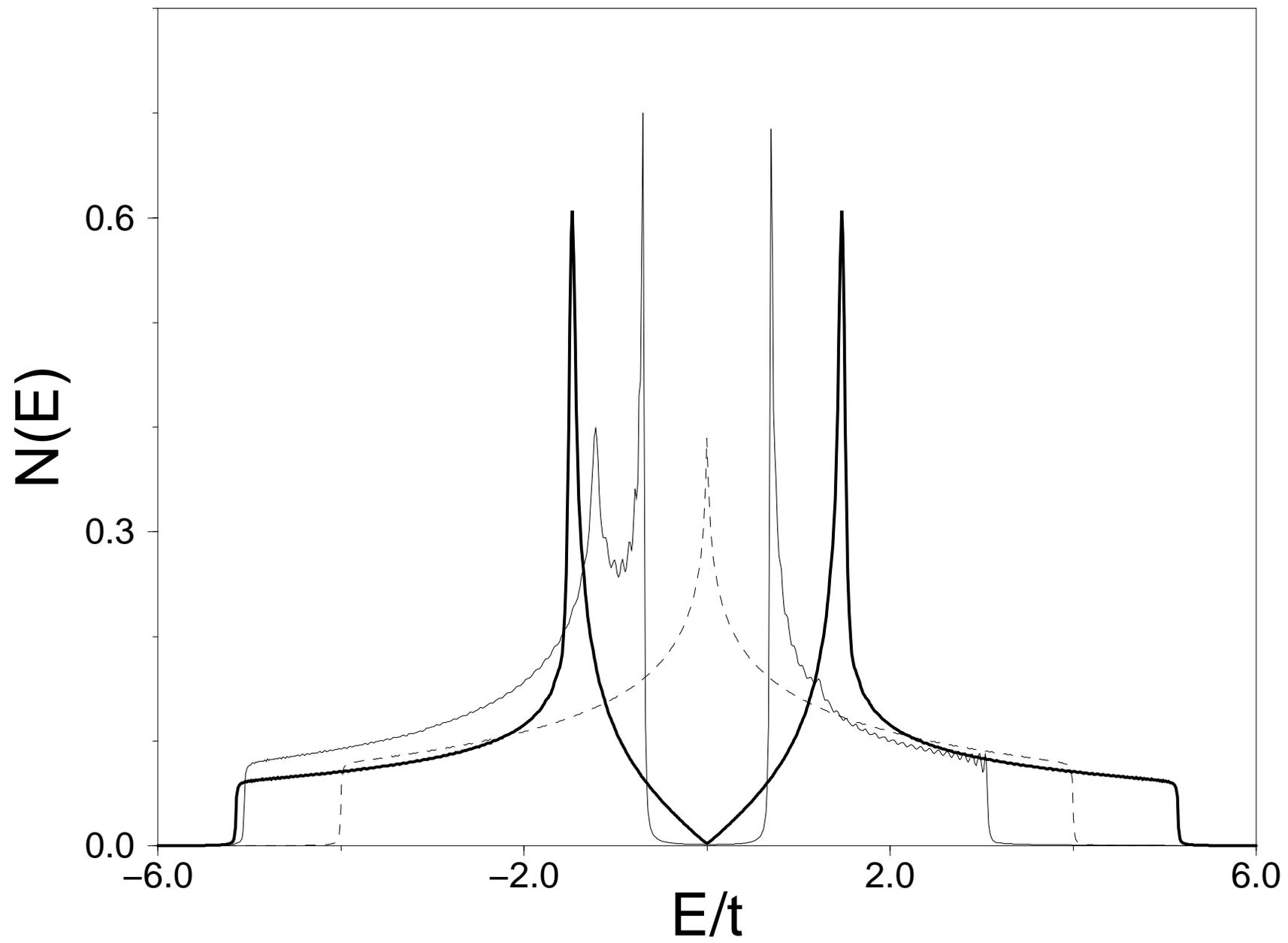
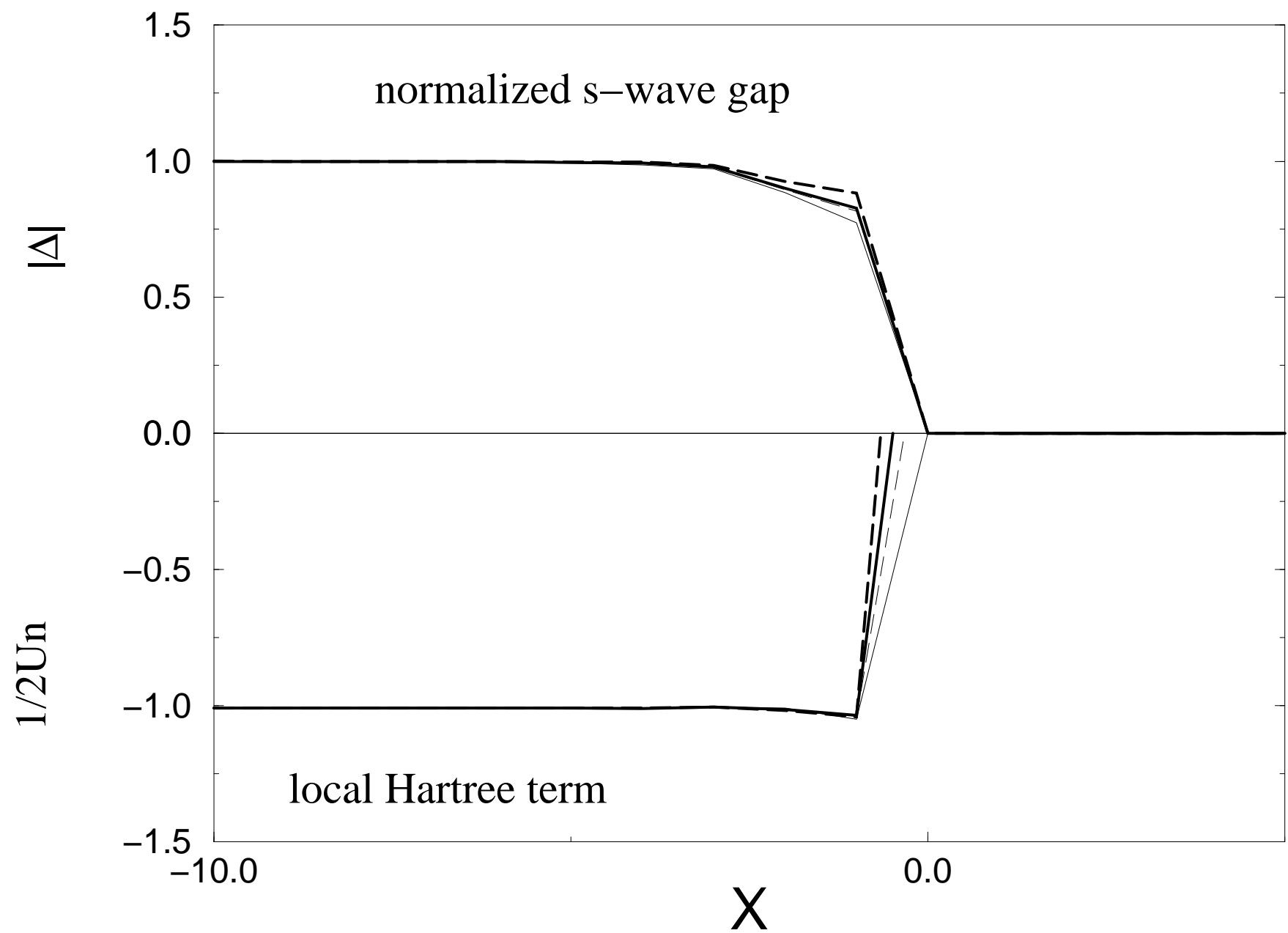
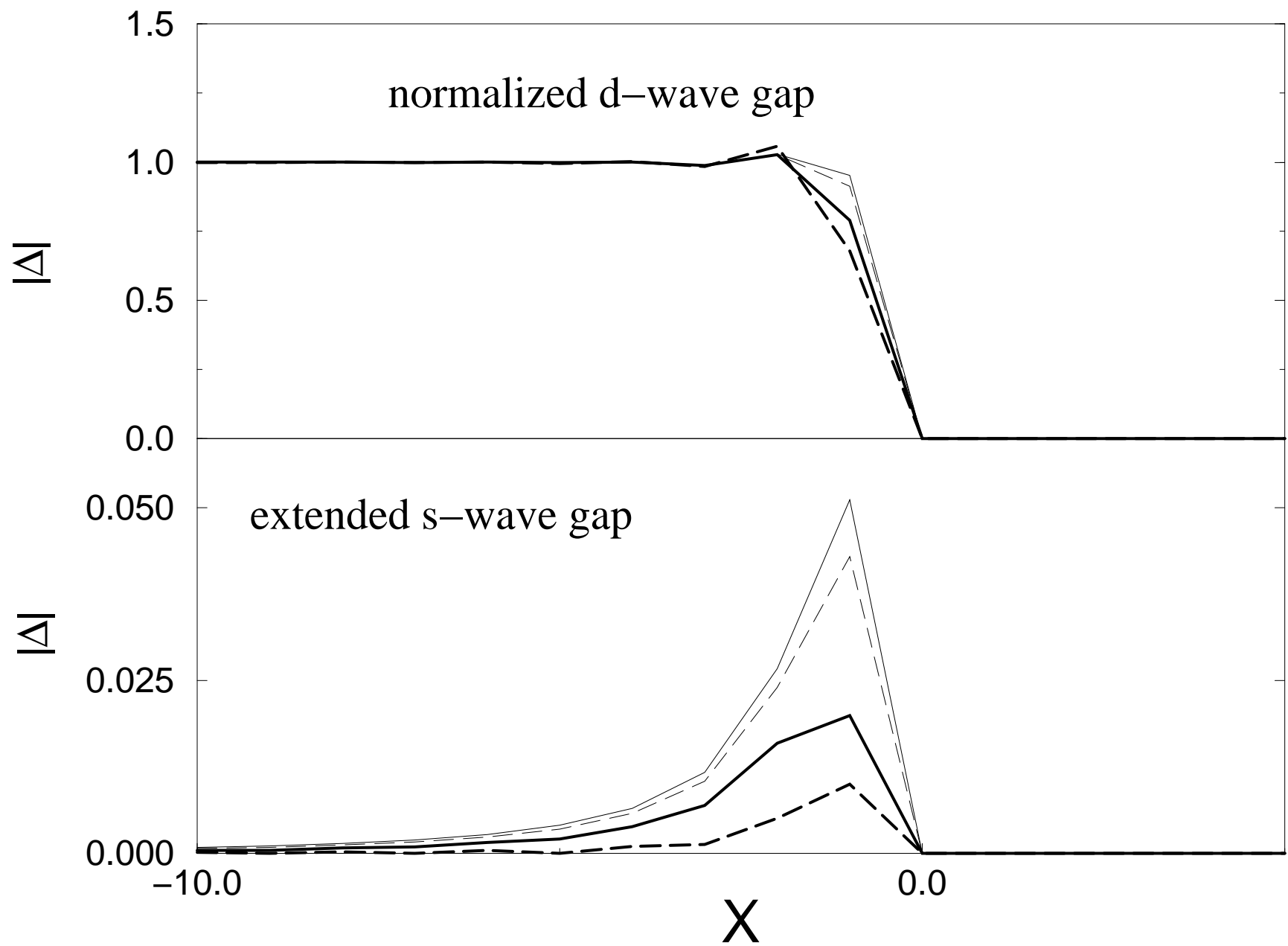
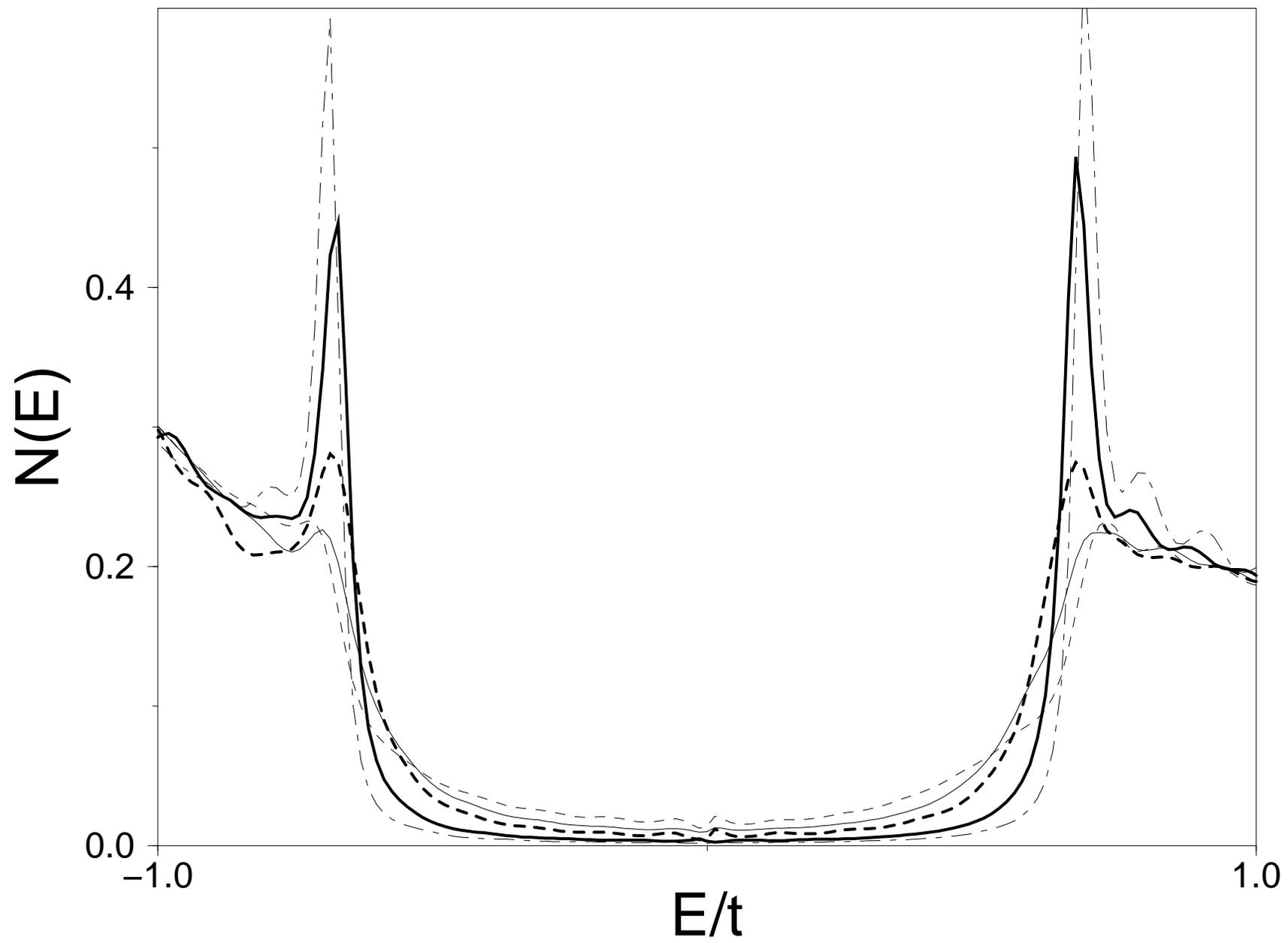
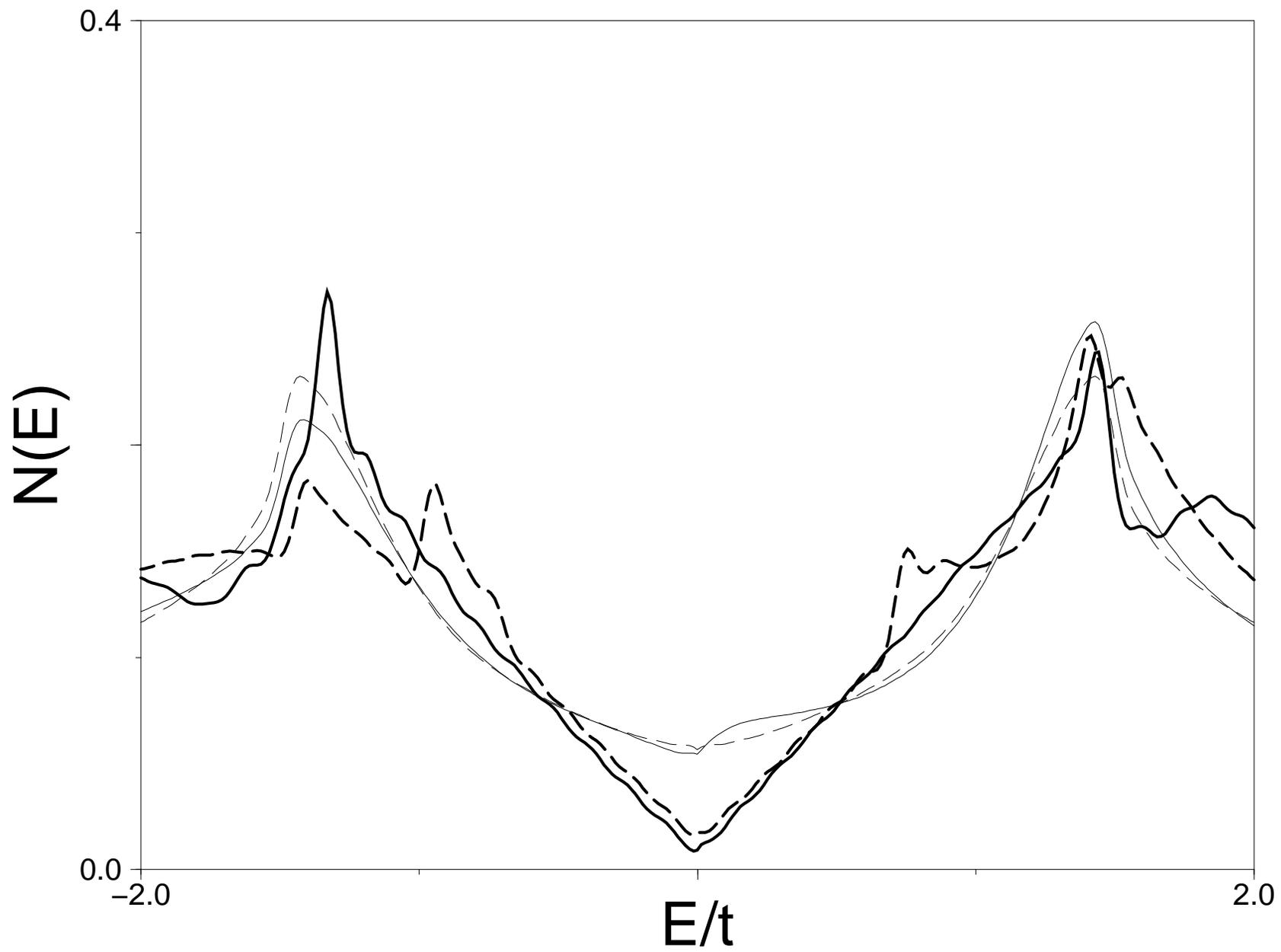
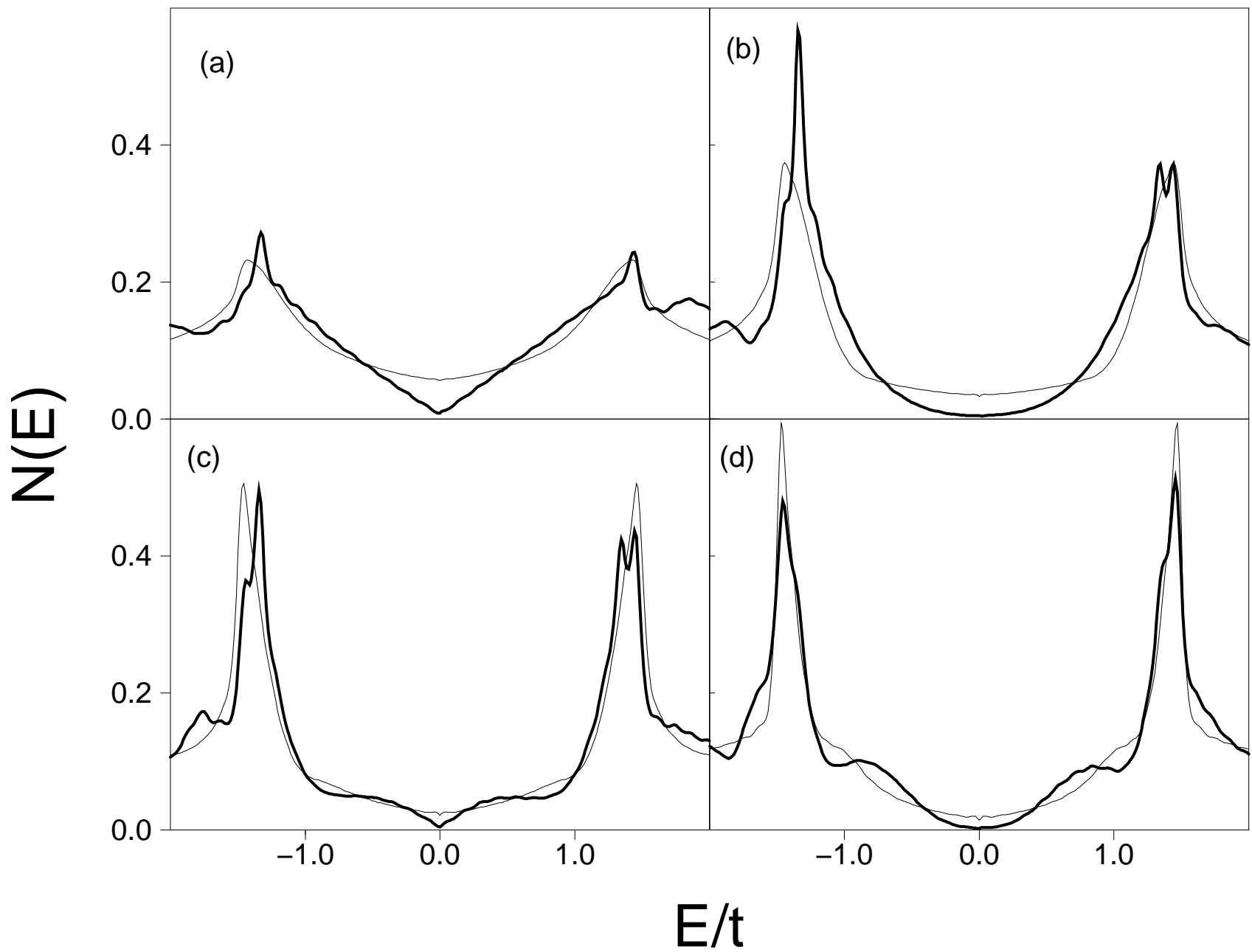
\end{document}